\documentclass[journal=jctcce,manuscript=article]{achemso}
\setkeys{acs}{articletitle = true}
\usepackage[version=3]{mhchem} 
\usepackage[T1]{fontenc}
\usepackage[utf8]{inputenc}

\usepackage{amssymb}
\usepackage{amsmath}
\usepackage{colortbl}
\usepackage{todonotes}

\usepackage{graphicx}
\usepackage{color}

\author{Mario Wolter}
\affiliation[TUBS]{Institute of Physical and Theoretical Chemistry, Technische Universität Braunschweig, Gaußstr.~17, 38106 Braunschweig, Germany}
\author{Moritz von Looz}
\affiliation[HUB]{Department of Computer Science, Humboldt-Universität zu Berlin, Unter den Linden 6, 10099 Berlin, Germany}
\altaffiliation{Now at Advanced Concepts Team, European Space Research and Technology Center, Noordwijk, The Netherlands}
\author{Henning Meyerhenke}
\affiliation[HUB]{Department of Computer Science, Humboldt-Universität zu Berlin, Unter den Linden 6, 10099 Berlin, Germany}
\email{meyerhenke@hu-berlin.de}
\author{\\Christoph R. Jacob}
\affiliation[TUBS]{Institute of Physical and Theoretical Chemistry, Technische Universität Braunschweig, Gaußstr.~17, 38106 Braunschweig, Germany}
\email{c.jacob@tu-braunschweig.de}

\title{Systematic partitioning of proteins for quantum-chemical fragmentation methods using graph algorithms}

\begin{document}

\newpage

\begin{abstract}
Quantum-chemical fragmentation methods offer an efficient approach for the treatment of large proteins, in particular if local 
target quantities such as protein--ligand interaction energies, enzymatic reaction energies, or spectroscopic properties of 
embedded chromophores are sought. However, the accuracy that is achievable for such local target quantities intricately 
depends on how the protein is partitioned into smaller fragments. While the commonly employed naïve approach of using 
fragments with a fixed size is widely used, it can result in large and unpredictable errors when varying the fragment size. Here, 
we present a systematic partitioning scheme that aims at minimizing the fragmentation error of a local target quantity for a given 
maximum fragment size. 
To this end, we construct a weighted graph representation of the protein, in which the amino acids constitute the nodes. These nodes
are connected by edges weighted with an estimate for the fragmentation error that is expected when cutting this edge. This 
allows us to employ graph partitioning algorithms provided by computer science to determine near-optimal partitions of the protein. 
We apply this scheme to a test set of six proteins representing various prototypical applications of quantum-chemical fragmentation
methods using a simplified molecular fractionation with conjugate caps (MFCC) approach with hydrogen caps. We show 
that our graph-based scheme consistently improves upon the naïve approach.
\end{abstract}

\vspace{4ex}
\begin{center}
  \textbf{Table of Contents Graphics} \\[2ex]

  \includegraphics[width=8.9cm]{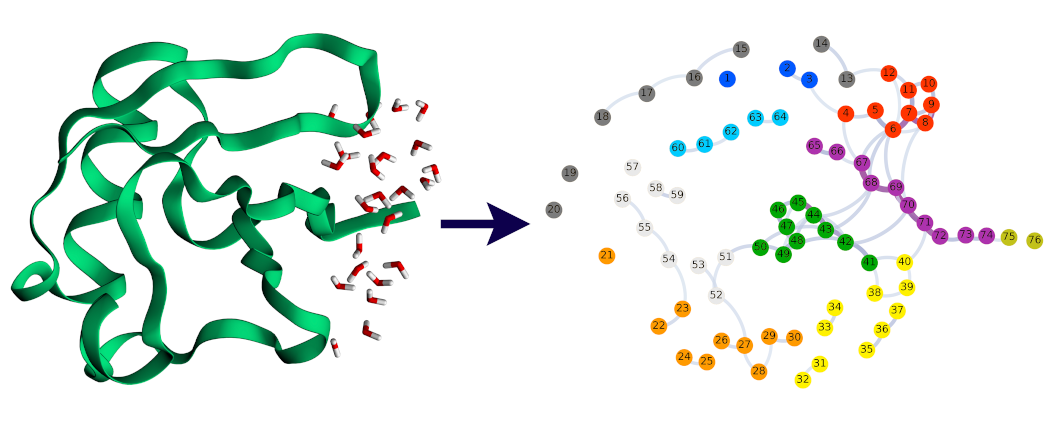}
\end{center}

\newpage

\section{Introduction}

Computer simulations have become an essential tool for investigating the chemical and physical mechanisms underlying the 
functionality of biomolecules \cite{karplus_development_2014,hollingsworth_molecular_2018}. With the development of classical 
force fields and increasingly powerful computational resources, structural investigation of biomolecules, comprised of up to hundred 
thousands of atoms, have become possible \cite{chung_switch-like_2019,venable_molecular_2019}. 
However, in many cases quantum-chemical methods are essential for accurately modeling biomolecular systems. Examples
include the accurate prediction of protein--ligand interaction energies \cite{ryde_ligand-binding_2016}, the calculation of 
spectroscopic properties of chromophores embedded in proteins \cite{knig_quantum_2012,bravaya_quantum_2012,
curutchet_quantum_2017}, and the modeling of enzymatic reactions \cite{romero-rivera_computational_2016,ahmadi_multiscale_2018}.

In the past decades, enormous progress has been made to increase the computational efficiency of quantum-chemical methods 
for large molecular systems. Efficient implementations nowadays allow for a full treatment of biomolecules containing hundreds up to 
thousands of atoms \cite{ochsenfeld_linear-scaling_2007,riplinger_natural_2013,robach_influence_2017}.  A particularly appealing 
strategy for the quantum-chemical treatment of large biomolecular systems is provided by fragmentation methods (for reviews, 
see, e.g., Refs.~\citenum{gordon_fragmentation_2012,raghavachari_accurate_2015,gordon_fragmentation:_2017}). 
Such methods partition a biomolecule into many smaller, possibly overlapping fragments, which are each treated individually. This generally 
leads to a natural linear scaling of the computational effort with the number of fragments, and thus significantly reduces the computational 
effort required for a treatment of the full biomolecular system. Moreover, the many independent calculations for the different fragments can 
easily be performed in an embarrassingly parallel fashion.

Numerous flavors of such fragmentation methods that are applicable to biomolecules have been developed. The original molecular fractionation 
with conjugate caps (MFCC) approach \cite{mfcc-1,mfcc-2,he_fragment_2014} partitions a protein into single amino acid residues, which are 
saturated with suitable capping groups, and performs separate quantum-chemical calculations for each resulting fragment. Such a treatment can 
be combined with the use of an embedding potential in the fragment calculations \cite{gomes_quantum-chemical_2012,goez_embedding_2018}. 
For instance, the MFCC method can be combined with an electrostatic embedding potential \cite{wang_electrostatically_2013,
jiang_electrostatic_2006}. The fragment molecular orbital (FMO) method \cite{fmo-1,fmo-2,fmo-review,tanaka_electron-correlated_2014,
fedorov_fragment_2017} employs a more sophisticated embedding potential in combination with a frozen-orbital treatment of the fragment
boundaries. One of us has previously developed an extension of the frozen-density embedding (FDE) scheme \cite{jacob_subsystem_2014,
wesolowski_frozen-density_2015} for the subsystem treatment of proteins, which employs the MFCC partitioning in combination with a 
density-based embedding potential (termed 3-FDE) \cite{3fde-2008}.

Fragmentation methods can be particularly useful if one is interested in properties that are local to a specific part of a biomolecular system.
In this case, they make it possible to selectively tune the accuracy of the local target quantities by employing a more accurate 
quantum-chemical description for fragments that are closer to the relevant region, while a less accurate description can be used for
distant fragments. Such a strategy can, for instance, be employed in the calculation of protein--ligand interaction energies, which demand
an accurate description of the protein's binding pocket, while for other parts of the protein the accuracy is less crucial \cite{karin-3fde,
asada_efficient_2012}. Another example is the calculation of spectroscopic properties of chromophores embedded in proteins. Here, 
fragmentation methods can be used to obtain an electron density of the protein, which can be included in the calculation of the 
chromophores' spectroscopic properties, e.g., via a frozen-density embedding potential \cite{goez_modeling_2014,goez_including_2017}. 
Again, in such a setup more accurate quantum-chemical methods could be used for fragments that are closer to the chromophores. 
We note that, in contrast to QM/MM methods \cite{senn_qm/mm_2009,morzan_spectroscopy_2018}, fragmentation methods generally 
maintain a quantum-chemical description for the whole protein.

A central issue in all fragmentation methods is the suitable choice of the fragments, i.e., the partitioning scheme used for decomposing 
the large biomolecular system into smaller subsystems. This choice requires a trade-off between efficiency and accuracy: In general, 
decreasing the size of the fragments will reduce the overall computational effort, but will at the same time increase the number of fragments
and thus also the error caused by the fragmentation. In the simplest and most common partitioning scheme, fragments consisting of a fixed 
number of amino acids along the backbone chain are used \cite{karin-3fde}. Antony and Grimme assessed such an approach for the calculation 
of protein--ligand interaction energies with MFCC and found a rather irregular dependence of the fragmentation error on the fragment size 
instead of a smooth convergence \cite{antony_fully_2012}. Thus, particularly for localized target quantities, the resulting fragmentation error 
can be highly dependent on the precise choice of the fragments. 

A number of partitioning schemes aim to address this issue with a more sophisticated definition of the relevant fragments. In the generalized 
MFCC scheme, overlapping fragments are defined based on chemical connectivity and spatial proximity such that relevant chemical interactions 
(e.g., hydrogen bonds between different chains) are included within the fragments \cite{mfcc-gmfcc,wang_electrostatically_2013}. Similarly, the
generalized energy-based fragmentation approach uses a distance cut-off when partitioning a protein into fragments \cite{li_generalized_2007,
li_generalized_2014}. The molecular tailoring approach (MTA) combines such a distance criterion with an algorithm for the iterative 
combination of fragments, which also accounts for chemical  connectivity information \cite{gadre_molecular_1994,babu_ab_2003,
sahu_molecular_2014}. However, all these schemes aim at the overall error in global properties of the full protein instead of the error 
in local target quantities such as protein--ligand interaction energies or spectroscopic properties.

Here, we present a systematic partitioning scheme that is tailored to the calculation of local target quantities, such as protein--ligand
interaction energies or local spectroscopic properties, with quantum-chemical fragmentation methods. Our partitioning scheme is constructed 
such that the expected fragmentation error in the target quantity is minimized, subject to constraints on the fragment size and/or the number of
fragments. To this end, we map the biomolecular system to a graph representation, in which the nodes represent amino acid residues
and the edges describe interactions between them. Each edge is weighted with an estimate of the error in the target quantity that is
expected when cutting this edge, i.e., when assigning the nodes connected by this edge to different fragments.

Using this representation, graph partitioning algorithms \cite{bulu_recent_2016} can be used to find a fragmentation that minimizes the 
\textit{edge cut weight}, (i.e., the sum of the weights of the edges between different fragments), which corresponds to the expected error 
in the local target quantity. 
We have previously put forward this idea and have presented suitable graph partitioning algorithms that include additional chemical 
constraints on the resulting partition \cite{von_looz_better_2016}. However, as the focus of this earlier work was on the algorithm engineering
aspects, it did not include any quantum-chemical calculations and only considered graphs that were constructed in an \textit{ad hoc} 
fashion. Here, we close this gap by applying our earlier ideas to actual quantum-chemical fragmentation calculations. We address the 
construction of suitable graph representations and carefully assess the accuracy of local target quantities with the resulting fragmentations.
We show that our graph-based, systematic partitioning scheme significantly improves the accuracy compared to the commonly used, simpler 
alternative schemes, while not increasing the overall computational effort.

This work is organized as follows. In Sect.~\ref{sec:method} we review the quantum-chemical fragmentation strategy employed 
here (Sect.~\ref{sec:frag-methods}), introduce measures for quantifying the fragmentation error (Sect.~\ref{sec:frag-error}), and 
present the methodology for mapping the fragmentation of proteins to a graph partitioning problem (Sects.~\ref{sec:protein-as-graph},
\ref{sec:edge-weights}, and~\ref{sec:dp-algorithm}). The computational details are given in Sect.~\ref{sec:comp-det}. This is
followed by a first test and assessment of our methodology for ubiquitin in Sect.~\ref{sec:ubq}. Finally, a larger test set of five
additional proteins is considered in Sect.~\ref{sec:proteins}. Conclusions are drawn and an outlook is given in Sect.~\ref{sec:conclusion}.

\section{Methodology}
\label{sec:method}

\subsection{Fragmentation methods for proteins}
\label{sec:frag-methods}

All quantum-chemical fragmentation methods have in common that they partition a large biomolecular system into smaller subsystems.
When considering the electron density, they commonly approximate the total system's electron density $\rho_\text{tot}(\boldsymbol{r})$ as
\begin{equation}
  \label{eq:rho-tot-mfcc}
  \rho_\text{tot}(\boldsymbol{r})  \approx \rho^\text{fragm}_\text{tot}(\boldsymbol{r})
     = \sum_i \rho_i(\boldsymbol{r}) - \sum_j \rho^\text{cap}_j(\boldsymbol{r}),
\end{equation}
where $\rho_i(\boldsymbol{r})$ is the electron density of the $i$-th fragment, in which dangling bonds have been saturated with suitable 
capping groups, and $\rho^\text{cap}_j(\boldsymbol{r})$ are the densities of capping molecules that serve to correct for these capping groups.
This fragmentation can be formulated in a more general way using the inclusion--exclusion principle \cite{richard_generalized_2012}.

The MFCC method \cite{mfcc-1,mfcc-2,he_fragment_2014} uses the above fragmentation of the electron density, and employs amino acids (or 
groups of amino acids) as fragments. These are cut at the peptide bonds, and --CO-CH$_3$ and --NH-CH$_3$ capping groups are introduced 
to preserve the polarity of the peptide bond.
For each cut peptide bond, these capping groups are then joined to an N-methyl-acetamide molecule, which is subtracted to correct for the 
introduced capping groups.  In the original MFCC method, quantum-chemical calculations are performed for the isolated fragments as well 
as cap molecules. Numerous other fragmentation methods are derived from the MFCC scheme \cite{gordon_fragmentation_2012,
raghavachari_accurate_2015}.

\begin{figure}
  \begin{center}
    \includegraphics[width=0.65\linewidth]{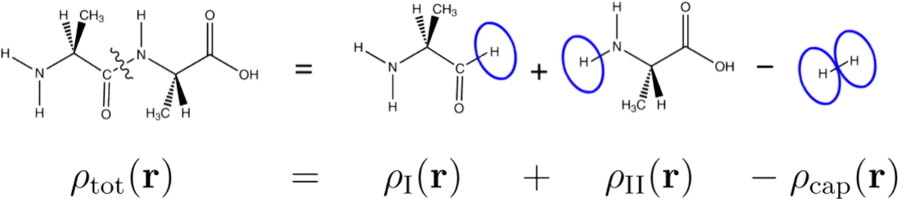}
  \end{center}
  \caption{Illustration of the fragmentation used in the simplified MFCC scheme with hydrogen caps that is employed throughout this work.}
  \label{fig:mfcc_h}
\end{figure}

Here, we aim to develop a partitioning scheme that is independent of the specific fragmentation method that is used. Therefore, in this work 
we consider only a simplified MFCC scheme in which hydrogen atoms are used as capping groups. The fragmentation and capping strategy used 
in this simplified scheme is illustrated in Fig.~\ref{fig:mfcc_h}. The use of hydrogen caps simplifies the mapping of the partitioning problem to a
graph, as there are no spurious interactions between the fragments and the cap molecules (see Sect.~\ref{sec:edge-weights} below). 
We note that by capping the peptide bond with only hydrogen atoms, the properties of the former peptide bond will be described inaccurately.
This will increase the overall errors introduced by the fragmentation, and will also introduce a stronger dependence of the total errors 
on the chosen partition. Thus, while the simplified MFCC scheme cannot be recommended for production calculations, it provides a good 
test case for the development and assessment of new algorithms for the fragmentation of proteins.

When considering individual amino acids as smallest possible fragments, a quantum-chemical fragmentation method requires to partition
the set consisting of the $N$ amino acid residues, $V = \{1,\dotsc,N\}$ into disjoint, non-empty subsets $V_1, \dotsc, V_k$ (corresponding
to fragments) such that each amino acid is assigned to exactly one subset, i.e., $\bigcup_{i} V_i = V$ and
$V_i \cap V_j = \emptyset$ for $i \neq j$. Here, we restrict ourselves to the problem formulation where each $V_i$ is a connected part
of the protein's main chain. Such a choice of fragments defines a partition $\Pi = \{V_1, \dotsc, V_k\}$ of the protein, and the accuracy 
of the results obtained with a quantum-chemical fragmentation method will depend on the chosen partition $\Pi$. Here, we want to 
determine this partition such that a suitably defined  error --- or an estimate for such an error ---  will be minimized.

The required computational effort of the quantum-chemical calculations will depend on the number of fragments $k$ and on the maximum 
size of the fragments $n_\text{max} = \max_i |V_i|$. These can be connected by defining an imbalance $\epsilon$ such that
\begin{equation}
  \label{eq:sizelimit}
  n_\text{max} \leq (1+ \epsilon) \left\lceil{\frac{N}{k}}\right\rceil, 
\end{equation}
where $\lceil x \rceil$ denotes the smallest integer greater than or equal to $x$. For $\epsilon = 0$ the partition is balanced. Here, our goal 
is to find an optimal partition along the protein's main chain for a given number of fragments $k$ and imbalance $\epsilon$, which implies 
a certain maximum size of the fragments $n_\text{max}$. For quantum-chemical fragmentation methods, it might be advantageous to allow 
for some imbalance of the fragment sizes and we choose $\epsilon = 0.33$ throughout this work.

\subsection{Definition of the fragmentation error}
\label{sec:frag-error}

In this work, we focus on the application of fragmentation methods for the calculation of local properties, e.g., protein--ligand interaction
energies, the properties of an enzymatic reaction center, or local spectroscopic properties. Thus, we are interested in the accurate 
description of the effect of the protein on a specific \textit{region of interest} (RoI). 

\begin{figure}
  \begin{center}
    \includegraphics[width=0.9\linewidth]{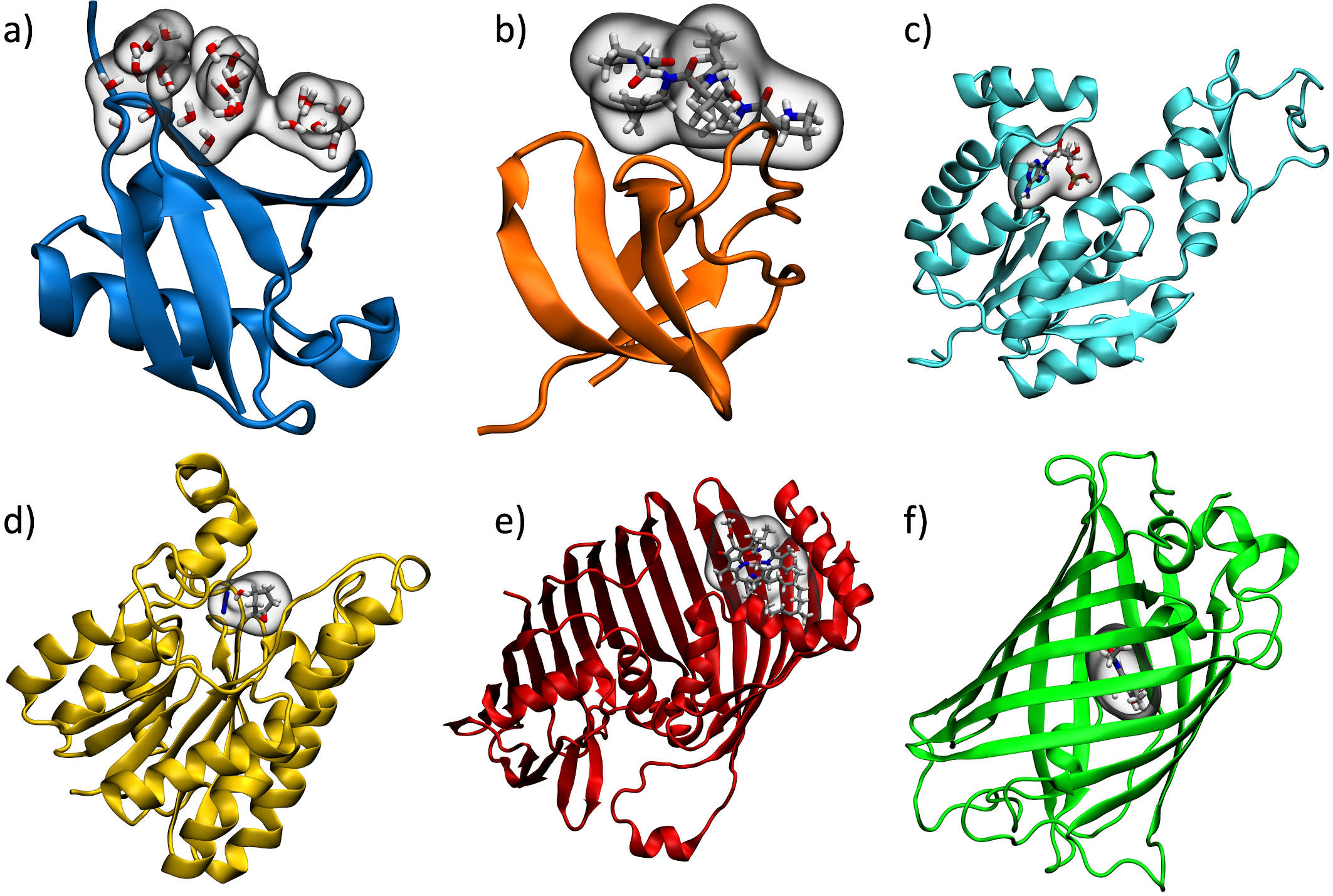}           
  \end{center}
  
  \caption{The test cases considered in this work for the calculation of local target properties with quantum-chemical fragmentation
                methods. The proteins are shown in a cartoon representation, whereas the respective binding partners, substrates, or 
                chromophores that define our region of interest (RoI) are shown in an atomistic representation and highlighted. 
                (a) Ubiquitin, (b) Sem5 SH3 Domain, (c) adenylate kinease, (d) halohydrin dehalogenase G (HheG), (e) Fenna-Matthews-Olson 
                complex (FMO), and (f) green fluorescent protein (GFP).}
  \label{fig:proteins_roi}
\end{figure}

Prototypical targets, which we will use as our test cases in the following, are shown in Fig.~\ref{fig:proteins_roi}. For protein--ligand 
interactions, we consider the interaction of ubiquitin with ligands or other proteins \cite{karin-3fde} (see Fig.~\ref{fig:proteins_roi}a) 
and the interaction of the \textit{Caenorhabditis elegans} adapter protein Sem5 Src homology~3 (SH3) domain with a peptoid 
inhibitor \cite{antony_fully_2012,nguyen_exploiting_1998} (see Fig.~\ref{fig:proteins_roi}b). 
As examples of enzymatic reaction centers, we consider the interaction of adenylate kinase with its substrate adenosine 
monophosphate \cite{antony_fully_2012} (see Fig.~\ref{fig:proteins_roi}c) as well as the interaction of halohydrin dehalogenase~G (HheG)
with a cyclohexene oxide, an azide and a water molecule in its binding pocket \cite{koopmeiners_hheg_2017} 
(see Fig.~\ref{fig:proteins_roi}d). Moreover, as test cases in which local spectroscopic properties are of interest, we consider a 
monomer taken from the Fenna-Matthews-Olsen complex (FMO) \cite{fenna_chlorophyll_1975,matthews_structure_1979,knig_protein_2013} 
(see Fig.~\ref{fig:proteins_roi}e) as well as the green fluorescent protein (GFP) \cite{zimmer_green_2002,bravaya_quantum_2012,
daday_chromophoreprotein_2015} (see Fig.~\ref{fig:proteins_roi}f). In these examples, the binding partner of the protein, the substrate 
of the enzyme, or the relevant chromophore define the respective RoIs, which are also highlighted in Fig.~\ref{fig:proteins_roi}. 
More details on each of our test cases as well as the choice of the RoI will be presented in Sects.~\ref{sec:ubq} 
and~\ref{sec:proteins}.

The fragmentation of the protein introduces an error in its total electron density, 
\begin{equation}
  \label{eq:dens_diff}
  \Delta \rho_\text{tot}(\boldsymbol{r}) = \rho^\text{super}_\text{tot}(\boldsymbol{r}) - \rho^\text{fragm}_\text{tot}(\boldsymbol{r}),
\end{equation}
where $\rho^\text{super}_\text{tot}(\boldsymbol{r})$ is the electron density obtained from a supermolecular calculation for the 
full protein. This error in the electron density translates to an error in the Coulomb potential within the RoI,
\begin{equation}
  \label{eq:coulomb_diff}
  \Delta V(\boldsymbol{r}) = \int \frac{ \Delta \rho_\text{tot}(\boldsymbol{r}') }{ |\boldsymbol{r} - \boldsymbol{r}'|}  {\rm d}^3r'
    = V^{\text{Coul}}_{\text{super}}(\boldsymbol{r}) - V^{\text{Coul}}_{\text{fragm}}(\boldsymbol{r}),
\end{equation}
where $V^{\text{Coul}}_{\text{super}}(\boldsymbol{r}) = \displaystyle\int \dfrac{\rho^\text{super}_\text{tot}(\boldsymbol{r}') }{ |\boldsymbol{r} 
- \boldsymbol{r}'|}  {\rm d}^3r'$ is the Coulomb potential obtained from a full, supermolecular calculation and 
$V^{\text{Coul}}_{\text{fragm}}(\boldsymbol{r}) = \displaystyle\int \dfrac{\rho^\text{fragm}_\text{tot}(\boldsymbol{r}') }{ |\boldsymbol{r} - \boldsymbol{r}'|}  
{\rm d}^3r'$ is the one obtained with the fragmentation method. The error in the Coulomb interaction energy between the protein electron density 
and the electron density of the region of interest is given by,
\begin{equation}
  \label{eq:def-error-signed}
  \Delta E = \int \rho_\text{RoI}(\boldsymbol{r}) \Delta V(\boldsymbol{r})  \, {\rm d}^3r,
\end{equation}
where $\rho_\text{RoI}(\boldsymbol{r})$ is the electron density of the ligand that constitutes the RoI, i.e., the error in the Coulomb 
potential is weighted with the electron density defining the RoI. When employing quantum-chemical fragmentation methods for the
calculation of electronic properties of the region of interest, this error in the Coulomb interaction energy $\Delta E$ should be kept as 
small as possible. We note that depending on the target quantity, other error measures might be needed. If protein--ligand interaction 
energies are of interest, also the errors in the electron--nuclear interaction energies should be included.

Of course, $\Delta E$ might be susceptible to error compensation between the contributions of different parts of the RoI. These will be
sensitive to the specific definition of the RoI and the choice of $\rho_\text{RoI}(\boldsymbol{r})$. Therefore, we define an absolute error
as
\begin{equation}
  \label{eq:def-error-abs}
  \Delta_{\text{abs}} = \int \rho_{\text{RoI}}(\mathbf{r}) \Big| \Delta V(\mathbf{r}) \Big| \, {\rm d}^3r.
\end{equation}
This absolute error is an upper bound for the error in the Coulomb interaction energy, i.e., $\Delta E \le \Delta_{\text{abs}}$. It can also
be expected that $\Delta_{\text{abs}}$ will control the error in spectroscopic properties that are localized in the RoI. In this work, our goal 
is to determine the partition $\Pi$ such that in a fragmentation calculation this absolute fragmentation error is minimized, subject to suitable 
constraints on the number of fragments and the imbalance.

\subsection{Protein fragmentation as graph partitioning problem}
\label{sec:protein-as-graph}

The problem of partitioning proteins for quantum-chemical fragmentation methods can be phrased as a version of the \emph{graph partitioning} 
problem considered in computer science \cite{bulu_recent_2016}.
A graph $G=(V,E)$ consists of a set of nodes (or vertices) $V$ and a set of edges 
$E$, with each edge connecting two nodes (we exclude self-loops).
A \emph{weighted graph} is a graph with a weight function $w(E)$, assigning an \emph{edge 
weight} to each edge.

\begin{figure}[ht]
  \begin{center}
    \includegraphics[width=0.85\linewidth]{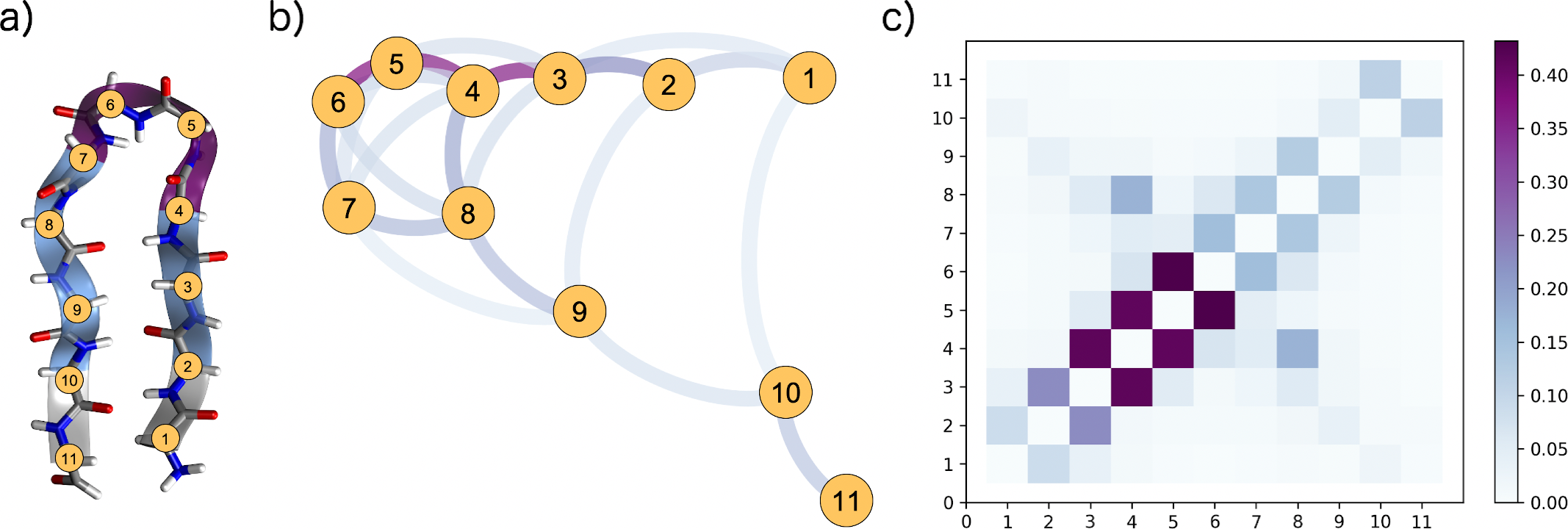}
  \end{center}
  \caption{Illustration of the representation of proteins by a weighted graph for the example of a $\beta$-hairpin structure. 
                (a) $\beta$-hairpin structure in atomistic and cartoon representation. (b) 2D representation of the corresponding
                weighted graph. The nodes are shown as yellow circles which are connected by edges. The weight of each edge
                is encoded in its color. (c) Representation of the symmetric matrix of the complete protein graph.}
  \label{fig:turn_mol_2d_graph}
\end{figure}

A protein consists of a linear main chain of amino acids, and the fragmentation schemes considered here use groups of amino acid 
residues as fragments. Therefore, the most straightforward way to map a protein to a graph is by using the individual amino acids 
as nodes $V$. Fig.~\ref{fig:turn_mol_2d_graph} illustrates this for the example of a $\beta$-hairpin. The eleven amino acids can be 
seen in an atomistic representation in Fig.~\ref{fig:turn_mol_2d_graph}a, whereas a graph representation in which the nodes are 
shown as yellow colored circles is presented in Fig.~\ref{fig:turn_mol_2d_graph}b.

Each edge $\{u,v\}$ in a graph connects two nodes $u$ and $v$, and when representing a protein as a graph, the edges connect the amino 
acids with each other. The most obvious choice is to connect the amino acids along the backbone chain with edges $\{u,u+1\}$, representing 
the bonded  interactions. This leads to a linear graph representing the primary structure of the protein. However, in proteins the main chain 
folds into a three-dimensional structure, stabilized by the non-bonded interactions between the individual amino acids. 
To reflect this, additional edges $\{u,v\}$ are introduced, representing non-bonded interactions between two amino acids. All edges are
assigned a suitable weight (see below). This is shown in Fig.~\ref{fig:turn_mol_2d_graph}b for the $\beta$-hairpin example, where the
weight is encoded in the color of the edges. 
A graph can be represented by a symmetric $N \times N$-matrix with the edge weights (or zero if no
edge is present between two nodes) as entries (see Fig.~\ref{fig:turn_mol_2d_graph}c). 
In the simplest approach, all amino acids in 
a protein can be connected by edges, the resulting graph is then called \emph{complete}.

With such a graph representation, defining a fragmentation of a protein becomes equivalent to partitioning the nodes of the graph into
subsets, i.e., to a graph partition $\Pi = \{V_1, \dotsc, V_k\}$ (cf.\ Sect.~\ref{sec:frag-methods}), possibly subject to additional constraints. 
Here, we restrict ourselves to partitions in which each subset is a connected sequence of nodes along the main chain of the protein, i.e.,
for a graph with ascending node IDs according to the node's main chain position, each subset must contain nodes within a continuous
ID range.

To determine this partition such that the expected fragmentation error is minimized, we define the weight $w(u,v)$ of the edge $\{u,v\}$ 
as the expected fragmentation error $\Delta_{\text{abs}}$ that arises if the two nodes (i.e., amino acids) $u$ and $v$ are assigned to 
different fragments. Thus, an estimate for the expected total fragmentation error that arises when using the partition $\Pi$ is given by
\begin{equation}
  \Delta_{\text{abs}}(\Pi,G) \approx C(\Pi,G) = \sum_{\substack{V_i,V_j \in \Pi \\ i \ne j}} \sum_{u \in V_i} \sum_{v \in V_j} w(u,v),
  \label{eq:error_edge_weights}
\end{equation}
that is, as the sum of the weights of all edges that need to be cut when partitioning the graph. Finding the general partition $\Pi_{k,\epsilon}$ 
(without the restriction of a continuous ID range per block) that minimizes this \textit{edge cut weight} $C(\Pi,G)$ for a given number of 
fragments $k$ and a given imbalance $\epsilon$ is known as \textit{graph partitioning} in computer science. The general problem is 
$\mathcal{NP}$-hard (and thus no efficient algorithm exists to solve it exactly), but fast algorithms have been developed that are able 
to find high-quality solutions \cite{bulu_recent_2016}. Previously, we have investigated different available algorithms, and have adapted 
those to the specific constraints that arise when considering quantum-chemical fragmentation methods \cite{von_looz_better_2016}. 
The restricted problem version we consider in this paper, in turn, can be solved exactly in polynomial time. Further details on the graph 
partitioning algorithm used here are presented below in Sect.~\ref{sec:dp-algorithm}.

\subsection{Calculation of the edge weights}
\label{sec:edge-weights}

For a given partition $\Pi = \{V_1, \dotsc, V_k\}$, our simplified MFCC scheme calculates the total electron density according 
to Eq.~\eqref{eq:rho-tot-mfcc}, where $\rho_i(\boldsymbol{r})$ corresponds to the electron density calculated for a fragment 
consisting of the amino acid residues in the fragments $V_i$ (suitably capped using hydrogen atoms as described above), 
whereas $\rho_j^\text{cap}(\boldsymbol{r})$ are the densities of the corresponding dihydrogen cap molecules.

To approximate the fragmentation error of a given partition, we consider a simplified two-body expansion in terms of single amino 
acid fragments, i.e.,
\begin{equation}
  \rho_i(\boldsymbol{r}) = \sum_{u \in V_i} \rho_u(\boldsymbol{r}) - \sum_{j} \rho_j^\text{cap}(\boldsymbol{r}) 
                                           + \sum_{u,v \in V_i} \Delta \rho_{uv}^{(2)}(\boldsymbol{r})
\end{equation}
with
\begin{equation}
  \Delta \rho_{uv}^{(2)}(\boldsymbol{r}) 
       = \rho_{uv}(\boldsymbol{r}) - \rho_{u}(\boldsymbol{r}) - \rho_{v}(\boldsymbol{r}) - \rho_{uv}^\text{cap}(\boldsymbol{r})
\end{equation}
for pairs of amino acids that are directly connected by a peptide bond and with
\begin{equation}
  \Delta \rho_{uv}^{(2)}(\boldsymbol{r}) = \rho_{uv}(\boldsymbol{r}) - \rho_{u}(\boldsymbol{r}) - \rho_{v}(\boldsymbol{r}) 
\end{equation}
for all other pairs of amino acids. Here, $\rho_{uv}(\boldsymbol{r})$ is the electron density of a fragment consisting of the suitably 
capped amino acid residues $u$ and $v$, $\rho_{u}(\boldsymbol{r})$ and $\rho_{v}(\boldsymbol{r})$ are the electron densities
of the suitably capped single amino acids residues $u$ and $v$, respectively, and $\rho_{uv}^\text{cap}(\boldsymbol{r})$ is the
cap molecule compensating for the caps introduced when cutting the peptide bonds between amino acids $u$ and $v$. We note
that the treatment of the cap molecules in this simplified two-body expansion is not fully consistent, as interactions between the
amino acid residues and the cap molecules are neglected (see Ref.~\citenum{richard_generalized_2012} for a formulation of the 
many-body expansion using the inclusion--exclusion principle that addresses this issue). However, for the simplified MFCC scheme 
using hydrogen caps considered here, we find that neglecting these interactions is justified (see Sect.~\ref{sec:ass-uqb-error}). 

Using the above two-body expansion in terms of single amino acids for both the supermolecular electron density 
$\rho^\text{super}_\text{tot}(\boldsymbol{r})$ and the fragment electron densities $\rho_i(\boldsymbol{r})$, we obtain
\begin{equation}
  \Delta\rho_\text{tot}(\boldsymbol{r}) 
     =  \sum_{\substack{V_i,V_j \in \Pi \\ i \ne j}} \sum_{u \in V_i} \sum_{v \in V_j}  \Delta \rho_{uv}^{(2)}(\boldsymbol{r})
\end{equation}
and
\begin{equation}
  \Delta V(\boldsymbol{r}) 
    =  \sum_{\substack{V_i,V_j \in \Pi \\ i \ne j}} \sum_{u \in V_i} \sum_{v \in V_j}  \Delta V_{uv}^{(2)}(\boldsymbol{r}),
\end{equation}
with $\Delta V_{uv}^{(2)}(\boldsymbol{r}) = \displaystyle\int \dfrac{\rho_{uv}^{(2)}(\boldsymbol{r}') }{ |\boldsymbol{r} - \boldsymbol{r}'|}  
{\rm d}^3r'$. Thus, the error in the Coulomb interaction energy [cf.\ Eq.~\eqref{eq:def-error-signed}] can be approximated as
\begin{equation}
  \label{eq:def-error-de}
  \Delta E \approx \Delta E^{(2)} 
    =  \sum_{\substack{V_i,V_j \in \Pi \\ i \ne j}} \sum_{u \in V_i} \sum_{v \in V_j}  
         \int \rho_\text{RoI}(\boldsymbol{r}) \Delta V_{uv}^{(2)}(\boldsymbol{r}) \, {\rm d}^3r,
\end{equation}
and the absolute fragmentation error [cf.\ Eq.~\eqref{eq:def-error-abs}] can be approximated as
\begin{equation}
\label{eq:def-error-dabs}
  \Delta_\text{abs} \approx \Delta_\text{abs}^{(2)} 
    =   \int \rho_\text{RoI}(\boldsymbol{r}) \bigg| \sum_{\substack{V_i,V_j \in \Pi \\ i \ne j}} \sum_{u \in V_i} \sum_{v \in V_j}  
         \Delta V_{uv}^{(2)}(\boldsymbol{r}) \bigg | \, {\rm d}^3r.
\end{equation}
For this two-body approximation of the absolute fragmentation error we can establish an upper bound,
\begin{equation}
\label{eq:def-error-dabs2}
  \Delta^{(2)}_\text{abs} \le  \sum_{\substack{V_i,V_j \in \Pi \\ i \ne j}} \sum_{u \in V_i} \sum_{v \in V_j}  
         \int \rho_\text{RoI}(\boldsymbol{r}) \Big|  \Delta V_{uv}^{(2)}(\boldsymbol{r}) \Big| \, {\rm d}^3r.
\end{equation}
By defining the edge weights of our graph representation as 
\begin{equation}
\label{eq:def-edge-weights}
  w(u,v) =  \int \rho_\text{RoI}(\boldsymbol{r}) \Big|  \Delta V_{uv}^{(2)}(\boldsymbol{r}) \Big| \, {\rm d}^3r,
\end{equation}
the edge cut weight $C(\Pi, G)$ thus provides an upper bound for the two-body approximation $\Delta_\text{abs}^{(2)}$ of the 
absolute fragmentation error $\Delta_\text{abs}$, which in turn provides an upper bound for the two-body approximation of the 
error in the Coulomb interaction energy, i.e., 
\begin{equation}
\label{eq:def-error-de2}
  \big| \Delta E^{(2)} \big| \le \Delta^{(2)}_\text{abs} \le C(\Pi, G).
\end{equation}
Note that this error bound cannot be expected to be tight. Thus, the edge cut weight $C(\Pi, G)$, which is minimized by graph 
partitioning algorithms, is not equal to the fragmentation error. Nevertheless, it can be expected that by reducing this upper bound,
also the fragmentation error will be reduced.

In Sect.~\ref{sec:ubq}, we will (a) assess the accuracy of the simplified two-body expansion for estimating the fragmentation error 
and (b) explore the application of graph partitioning algorithms for systematically reducing the fragmentation error in our simplified
MFCC scheme.

\subsection{Dynamic Programming Graph Partitioning Algorithm}
\label{sec:dp-algorithm}

Our restricted version of the graph partitioning problem, in which we search for continuous fragments along the main chain 
(cf.\ Section~\ref{sec:protein-as-graph}), leads to the following formulation: Given a graph $G=(V,E)$ with ascending node IDs 
according to the node's main chain position, an integer $k$ and a maximum imbalance $\epsilon$, find a $k$-partition $\Pi_{k,\epsilon}$
which respects the balance constraints, minimizes the edge cut weight $C(\Pi,G)$ and satisfies $v_j \in V_i \wedge v_j+l \in V_i 
\rightarrow v_j+1 \in V_i$ for all $l \in \mathbb{N}^+, 1 \leq j < n, 1 \leq i \leq k$ (\textit{continuous node constraint}). 

Finding fragments with continuous node IDs is equivalent to finding a set of $k-1$ \emph{delimiter nodes} 
$v_{d_1}, v_{d_2}, ... v_{d_{k-1}}$ that separate the fragments. More precisely, delimiter node $v_{d_j}$
belongs to fragment $j$, $1 \leq j \leq k-1$. Consider the delimiter nodes in ascending order.
Given the node $v_{d_2}$, the optimal placement of node $v_{d_1}$ only depends on edges among nodes $u < v_{d_2}$,
since all edges $\{u,v\}$ from nodes $u < v_{d_2}$ to nodes $v > v_{d_2}$ are cut no matter where $v_{d_1}$ is placed.
Placing node $v_{d_2}$ thus induces an optimal placement for $v_{d_1}$, using only information from edges to nodes $u < v_{d_2}$.
With this dependency of the positions of $v_{d_1}$ and $v_{d_2}$, placing node $v_{d_3}$ similarly induces an optimal choice for 
$v_{d_2}$ and $v_{d_1}$, using only information from nodes smaller than $v_{d_3}$. The same argument can be continued inductively 
for nodes $v_{d_4} \dots v_{d_k}$.

This insight is used by our algorithm, which is based on the dynamic programming paradigm.
It iteratively computes the optimal placement of $v_{d_{j-1}}$ for all possible values of $v_{d_{j}}$. 
Finding the optimal placements of $v_{d_1}, \dots v_{d_{j-1}}$ given a delimiter $v_{d_{j}}$ at node $i$ is equivalent to 
the subproblem of partitioning the first $i$ nodes into $j$ fragments, for increasing values of $i$ and $j$. 
When $N$ nodes and $k$ fragments are reached, the desired global solution is found. For more details,
we refer to our previous paper \cite{von_looz_better_2016}.

\subsection{Computational Details}
\label{sec:comp-det}

All quantum-chemical calculations were performed using the Amsterdam Density Functional (ADF) package \cite{chem-with-adf, adf-scm}
using density-functional theory (DFT) with the BP86 exchange--correlation functional \cite{B88,Perdew86} and a Slater-type
DZP basis set \cite{van_lenthe_optimized_2003}. The complete workflow for the construction of the graph representation as well as for
our quantum-chemical fragmentation calculations using the simplified MFCC scheme has been implemented in the PyADF scripting 
framework \cite{pyadf-2011}. PyADF Version 0.97, which can be used to reproduce all calculations presented here, is available under
the GPL v3 license at Ref.~\citenum{pyadf-0-97}.

In the construction of the graph representation and in the underlying two-body approximation, we included only pairs of amino acids with a 
maximum distance of five amino acid units along the main chain or with a maximum distance of 2.5 {\AA} through space. An assessment of
the accuracy of applying such cut-offs is presented in the Supporting Information (see Figs.~S3 and~S4). The edge weights were calculated 
according to Eq.~\eqref{eq:def-edge-weights} from DFT calculations for the individual amino acids as well as the corresponding pairs. The 
integration was performed using ADF's integration grid for the respective RoI. 

The graphs were then exported to the Chaco/METIS format \cite{karypis_metis._2013} to enable the use of standard graph partitioning 
algorithms available in a dedicated fork of the NetworKit package \cite{staudt_networkit:_2016}
which implements the DP algorithm described above. The resulting partition was then imported into PyADF in order to perform the quantum-chemical fragmentation calculation. 

Details on the protein structures used as our test cases are given in Sects.~\ref{sec:ubq} and~\ref{sec:proteins}. All protein crystal 
structures were saturated with hydrogen atoms using Openbabel \cite{oboyle_open_2011,openbabel-241} to achieve a neutral 
protonation state for all amino acids. 3D visualizations of the protein structures were created with VMD \cite{vmd} and its build-in 
tools. All plots have been prepared using Python and Matplotlib \cite{hunter_matplotlib:_2007,matplotlib-3-2-1} in Jupyter notebooks.
Graph equilibrations and visualizations were performed with gephi \cite{gephi-1,gephi}.

A data set containing molecular structures of all proteins (including hydrogen atoms) and the corresponding RoIs, the constructed graph 
representation, the resulting partitions, raw data for all the figures as well as PyADF input files and Jupyter notebooks used for data analysis 
and for producing all figures is available in the Zenodo repository at Ref.~\citenum{dataset-graph-partition}.

\section{Initial test case: Ubiquitin}
\label{sec:ubq}

As an initial test case, we employ the small protein ubiquitin, which has been used previously as a test case for the 3-FDE 
scheme \cite{3fde-2008,karin-3fde}. Ubiquitin (PDB code: 1UBQ \cite{1ubq}) consists of 76 amino acids, which still allows for a 
supermolecular DFT calculation that can be used as reference. As the crystal structure does not come with a specific binding 
partner, a cluster of water molecules placed in the face region of ubiquitin was used as RoI \cite{karin-3fde}. To build this water 
cluster, the whole protein was solvated with TIP3P water, the coordinates of the solvent molecules were energy minimized for 
100~steps and a short molecular dynamics simulation of 20~ps was performed with VMD and NAMD \cite{namd,namd-1} 
using a classical force field (Charmm27 \cite{charmm-1,charmm-2}). After this procedure the 24 water molecules within a 
distance of 5~{\AA} around the residues Leu8, Ile44 and Val70 were assigned to our RoI (see Fig.~\ref{fig:proteins_roi}a).

\subsection{Assessment of error estimation}
\label{sec:ass-uqb-error}

First, we assess the different measures of the fragmentation error defined in Sect.~\ref{sec:frag-error} as well as the corresponding
error estimates introduced in Sect.~\ref{sec:edge-weights} for the ubiquitin test case. 
The most straightforward way to partition a protein into small subsystems is to use a naïve partitioning algorithm, in which the protein is 
partitioned into fragments containing a repeating number of amino acids along the backbone chain \cite{karin-3fde, antony_fully_2012}. 
Thus, this naïve partition is defined by the chosen number of amino acids per fragment $n_\text{max}$. We note that in this case, the last 
fragment at the end of the peptide chain might be smaller. Therefore, for $n_\text{max} \ge N/2$, where $N$ is the total number of amino 
acids, the protein is always partitioned into two fragments and the only difference is the position of the cut along the backbone chain.

\begin{figure}[ht]
	\begin{center}
		\includegraphics[width=0.75\linewidth]{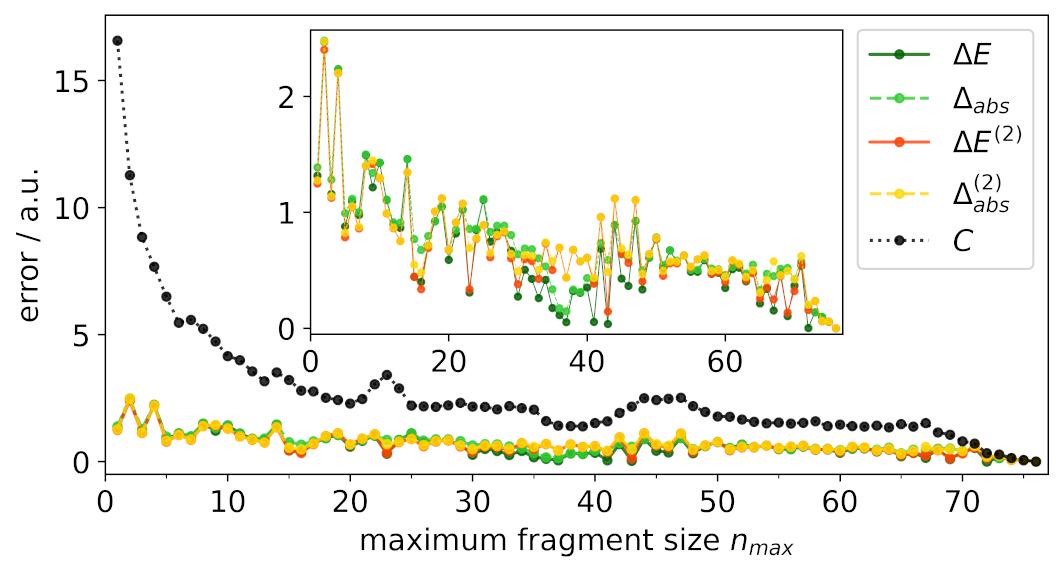}
		\caption{Comparison of fragmentation error measures $\Delta E$ (solid dark green line), $\Delta_\text{abs}$ (dashed green line)
		             and error estimates $\Delta E^{(2)}$ (solid red line), $\Delta_\text{abs}^{(2)}$ (dashed yellow line), $C$ (dotted black line)
		             for the ubiquitin test case using naïve partitions with different maximum number of amino acids per fragment $n_\text{max}$. 
		             The inset shows the same data but excludes $C$.}
		\label{fig:ubi_naïve_err_aspf}
	\end{center}
\end{figure}

In Fig.~\ref{fig:ubi_naïve_err_aspf}, the different fragmentation error measures are plotted for such a naïve partition with an 
increasing number $n_\text{max}$ of amino acids per fragment. Here, the error in the Coulomb interaction energy $\Delta E$ 
[cf.~Eq.~\eqref{eq:def-error-signed}] (solid dark green line) and the absolute error $\Delta_\text{abs}$ [cf.~Eq.~\eqref{eq:def-error-abs}] 
(dashed green line) have been calculated using the difference of the electron density and Coulomb potential obtained from a full, 
supermolecular DFT calculation of ubiquitin and from the respective simplified MFCC calculations. The electron density $\rho_\text{RoI}$ 
has been obtained from a DFT calculation of the 24 water molecules assigned to the RoI. Both error measures are in rather good agreement 
and show a similar trend. As expected, the absolute error $\Delta_\text{abs}$ is always larger than the error in the Coulomb interaction 
energy $\Delta E$, because in the latter positive and negative deviations of the Coulomb potential from the supermolecular reference 
can cancel each other.

Independent of the error measure, the graphs do not show a consistent decrease of the fragmentation error as the fragment size 
$n_\text{max}$ is increased. Instead, there are rather large fluctuations in the fragmentation error. Thus, the number of cuts between 
the fragments is not decisive for the accuracy of the fragmentation, but the main impact seems to be given by the specific positions 
and distributions of the cuts between the fragments.

The corresponding error estimates $\Delta E^{(2)}$ [cf.\ Eq.~\eqref{eq:def-error-de}] and $\Delta_\text{abs}^{(2)}$ [cf.\ Eq.~\eqref{eq:def-error-dabs}] 
are included in Fig.~\ref{fig:ubi_naïve_err_aspf} as solid red and dashed yellow line, respectively. Altogether, these are in rather good
agreement with the corresponding error measures. There are, however, some partitions (e.g., for fragments sizes $n_\text{max}$ 
between 35 and 40) where larger differences are found. Nevertheless, the overall comparison for ubiquitin confirms that the two-body 
approximation for the electron density, on which our error estimates are based, is sufficiently accurate and that neglecting the interaction 
between the hydrogen caps and the fragments is justified for our simplified MFCC scheme. This is confirmed by results for a second test 
case (Sem5 SH3 domain, see Section~\ref{sec:proteins}), which are shown in the Supporting Informations (see Figs.~S1 and~S2).

Finally, the edge cut weight $C(\Pi,G)$ [cf.\ Eq.~\eqref{eq:error_edge_weights}] calculated using the edge weights as defined in 
Eq.~\eqref{eq:def-edge-weights} is also shown Fig.~\ref{fig:ubi_naïve_err_aspf} (dotted black line). This edge cut weight constitutes
an upper bound for the absolute error estimate $\Delta_\text{abs}^{(2)}$. For our test case, it is always significantly larger than
the corresponding error estimates, but roughly follows a similar trend, i.e., it decreases as the fragment size $n_\text{max}$ is
increased. Whether the edge cut weight is nevertheless useful for systematically reducing the fragmentation error will be explored
in the following.

\subsection{Application of graph partitioning algorithms}
\label{sec:ass-ubq-graph}

Graph partitioning algorithms, such as the exact dynamic programming (DP) algorithm described in Sect.~\ref{sec:dp-algorithm}, can be 
applied to find a partition of the graph representation of a protein that (approximately) minimizes the edge cut weight $C(\Pi, G)$. 
With the definition of the edge weights given by Eq.~\eqref{eq:error_edge_weights}, this partition will thus (approximately) minimize 
an upper bound for our estimate $\Delta_\text{abs}^{(2)}$ of the absolute error $\Delta_\text{abs}$.

\begin{figure}
	\begin{center}
		\includegraphics[width=0.9\linewidth]{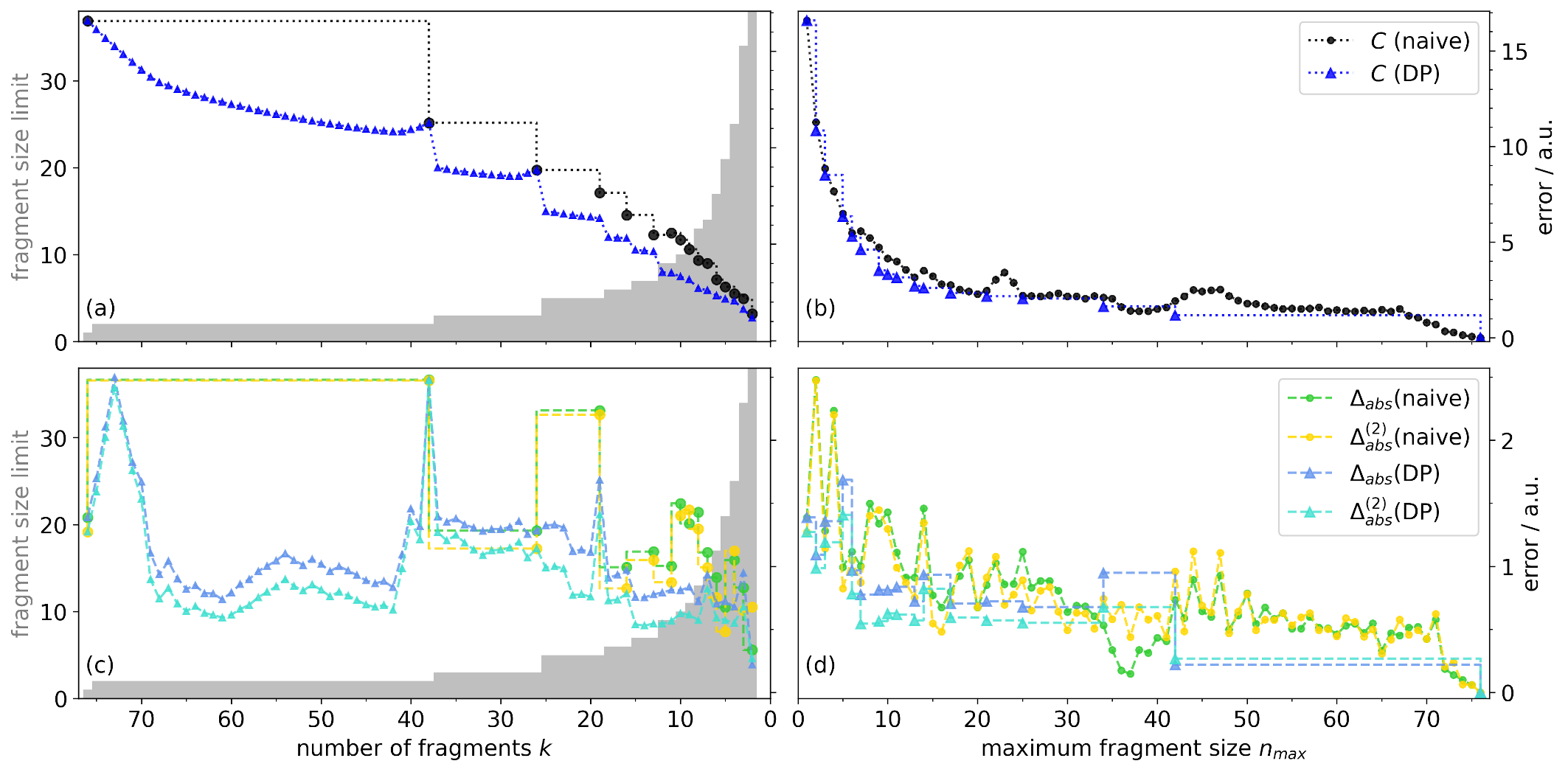}
	\end{center}

	\caption{Comparison of the partitions obtained by the DP algorithm (triangles) and the corresponding naïve partitions (circles).
	              (a,b) Comparison of the edge cut weight. (c,d) Comparison of the absolute error $\Delta_\text{abs}$ (green circles
	              and blue triangles) and the corresponding error estimate $\Delta_\text{abs}^{(2)}$ (yellow circles and cyan triangles). 
	              In (a) and (c) the horizontal axis shows the number of fragments $k$. The corresponding fragment size limit
	              $n_\text{max}$ applicable in the DP algorithm is shown as gray bars. 
	              In (b) and (d), the horizontal axis shows the maximum number of amino acids per fragment $n_\text{max}$. 
	              See text for details.\\[1ex]}
	\label{fig:ubi_naïve_dp}
\end{figure}

Fig.~\ref{fig:ubi_naïve_dp}a shows the edge cut weight $C$ of the partition obtained with the DP algorithm and an imbalance 
of $\epsilon = 0.33$ (blue triangles) as a function of the number of subsystems~$k$, which is used as input parameter in the DP
algorithm. The maximum number of amino acids per fragment $n_\text{max}$ corresponding to a given $k$ according to 
Eq.~\eqref{eq:sizelimit} is included as gray bars. For comparison, the edge cut weight $C$ obtained with the naïve partitioning 
scheme is also included (black circles). Note that a naïve partition is not available for arbitrary numbers of fragments $k$, but only 
for $k = \lceil\frac{N}{n_\text{max}}\rceil$ with $n_\text{max} = 1,2,\dotsc$ (black circles). For values of $k$ between those for 
which a naïve partition is available, $n_\text{max} = \lceil\frac{N}{k}\rceil$ is used (black dotted line), i.e., the naïve partition using 
a smaller number of fragments but the same maximum fragment size. Note that $n_\text{max}$ in the naïve partition and in the 
DP algorithm do not agree because the naïve partition implies a balanced partition ($\epsilon = 0$), whereas $\epsilon = 0.33$ 
is used in the DP algorithm.

Except for rather large numbers of fragments, the DP algorithm is able to find partitions with a lower $C$ for almost all $k$. 
For $k=76$, $k=38$, and $k=26$ the DP algorithm results in the corresponding naïve partition. In these cases, the choice 
of an imbalance of $\epsilon = 0.33$ requires that all fragments have the same size ($n_\text{max} = 1,2,3$, respectively), 
which restricts the DP algorithm to the naïve partition. Thus, the partition found by the DP algorithm is always as least as
good as the naïve partition, and it improves upon the latter if sufficient degrees of freedom are available. The improvement in 
the edge cut weight $C$ brought about by the DP algorithm is particularly obvious for $k$ between 5 and 20 fragments, 
which correspond to fragment sizes between 5 and 21 that would be feasible in large-scale applications of quantum-chemical 
fragmentation methods.

However, for the computational efficiency in quantum-chemical fragmentation methods, the size of the individual fragments is 
key, while the number of fragments is less important, in particular if the fragment calculations can be executed in parallel.
Therefore, Fig.~\ref{fig:ubi_naïve_dp}b shows an alternative presentation of the same data, in which $n_\text{max}$ is used as 
parameter on the horizontal axis. This way, the horizontal axis directly corresponds to the computational effort for the individual 
quantum-chemical calculations.

Here, a naïve partition is available for each integer $n_\text{max} \le N$. For the DP algorithm, different numbers of fragments $k$ can 
result in the same maximum fragment size $n_\text{max}$ according to Eq.~\eqref{eq:sizelimit}. Therefore, we apply the DP algorithm 
for all fragment numbers $k$ that result in the same $n_\text{max}$ and choose the partition $\Pi_{n_\text{max},\epsilon}$ that minimizes 
the edge cut weight, i.e.,
\begin{equation}
	C(\Pi_{{n}_\text{max},\epsilon}, G) = \min_{k \rightarrow n_\text{max}} C(\Pi_{k,\epsilon}, G).
\end{equation}
It is possible that no $k$ results in a given integer $n_\text{max}$ if the chosen maximum fragment size becomes large. In this case,
the largest integer $k$ that results in an $n_\text{max}$ that is smaller than or equal to the chosen $n_\text{max}$ is used and indicated
by the blue dotted line.

For small maximum fragment size up to four amino acids per fragment, there is no or only little freedom to optimize the partition and
the DP algorithm does not find better partitions than the naïve approach. Starting at a maximum fragment size of five, the DP algorithm
consistently outperforms the naïve approach and is able to find a partition with a lower edge cut weight.

The edge cut weight $C$ is only an upper bound for the error estimate $\Delta_\text{abs}^{(2)}$, which approximates the absolute fragmentation
error $\Delta_\text{abs}$. Therefore, we need to compare whether the reduction of $C$ for the partition obtained with the DP algorithm 
also corresponds to a reduction of $\Delta_\text{abs}^{(2)}$ and $\Delta_\text{abs}$. Fig.~\ref{fig:ubi_naïve_dp}c and~d compare these for
the partition obtained from the DP algorithm (blue and cyan triangles) and the naïve partition (green and yellow circles) as a function 
of the number of fragments $k$ and of the maximum number of amino acids per fragment $n_\text{max}$, respectively.

First, we note that the error estimate $\Delta_\text{abs}^{(2)}$ follows same trends as $\Delta_\text{abs}$. However, some differences 
appear, which are larger for the partitions from the DP algorithm than for the naïve partition, for instance for 40--70 fragments and for 
20--25 fragments in Fig.~\ref{fig:ubi_naïve_dp}c and for a maximum fragment size between 35 and 40 in Fig.~\ref{fig:ubi_naïve_dp}d. 
Nevertheless, the two-body error estimate appears to be a good approximate to the absolute error that is sufficiently accurate for our 
purposes.

In Fig.~\ref{fig:ubi_naïve_dp}c, the region that is most relevant for applications of quantum-chemical fragmentation methods is between
5 and 20 fragments, corresponding to maximum fragment sizes between 5 and 21 amino acids. Here, the DP algorithm reduces the 
absolute fragmentation error by about half compared to the naïve partition. In Fig.~\ref{fig:ubi_naïve_dp}d, we find that the DP algorithm 
consistently reduces the absolute fragmentation error for fragments of up to 30 amino acids. For very small fragments there is only
little flexibility in the choice of partition and thus for fragments sizes 1, 3, and 4 the DP algorithm cannot improve upon the naïve partition.
For fragment sizes of 15 and 16, the absolute fragmentation error of the partition found by the DP algorithm is slightly larger than for
the naïve partition, but in these cases also the naïve partition gives a comparably small fragmentation error, probably because of 
fortunate error cancellation. For fragment sizes of 33 to 40 amino acids, the error estimate $\Delta_\text{abs}^{(2)}$ is comparable for
the DP and naïve partition. However, in these cases the deviation of this error estimate from the actual absolute error is particularly
large and thus the absolute error is significantly larger for the DP algorithm than for the naïve partition. Note again that the DP algorithm 
does not optimize $\Delta_\text{abs}^{(2)}$ directly, but the edge cut weight $C$, which is an upper bound. 

Despite some discrepancies, overall the DP algorithm improves significantly compared to the naïve partition. In the relevant region, it can
reduce the absolute fragmentation error $\Delta_\text{abs}$ by up to a factor of two. Moreover, the DP algorithm leads to partitions 
for which the absolute fragmentation error tends to be reduced more consistently when increasing the maximum fragment size, whereas
for the naïve partition the absolute fragmentation error can show rather large oscillations with the maximum fragment size.

\begin{figure}
	\begin{center}
		\includegraphics[width=0.9\linewidth]{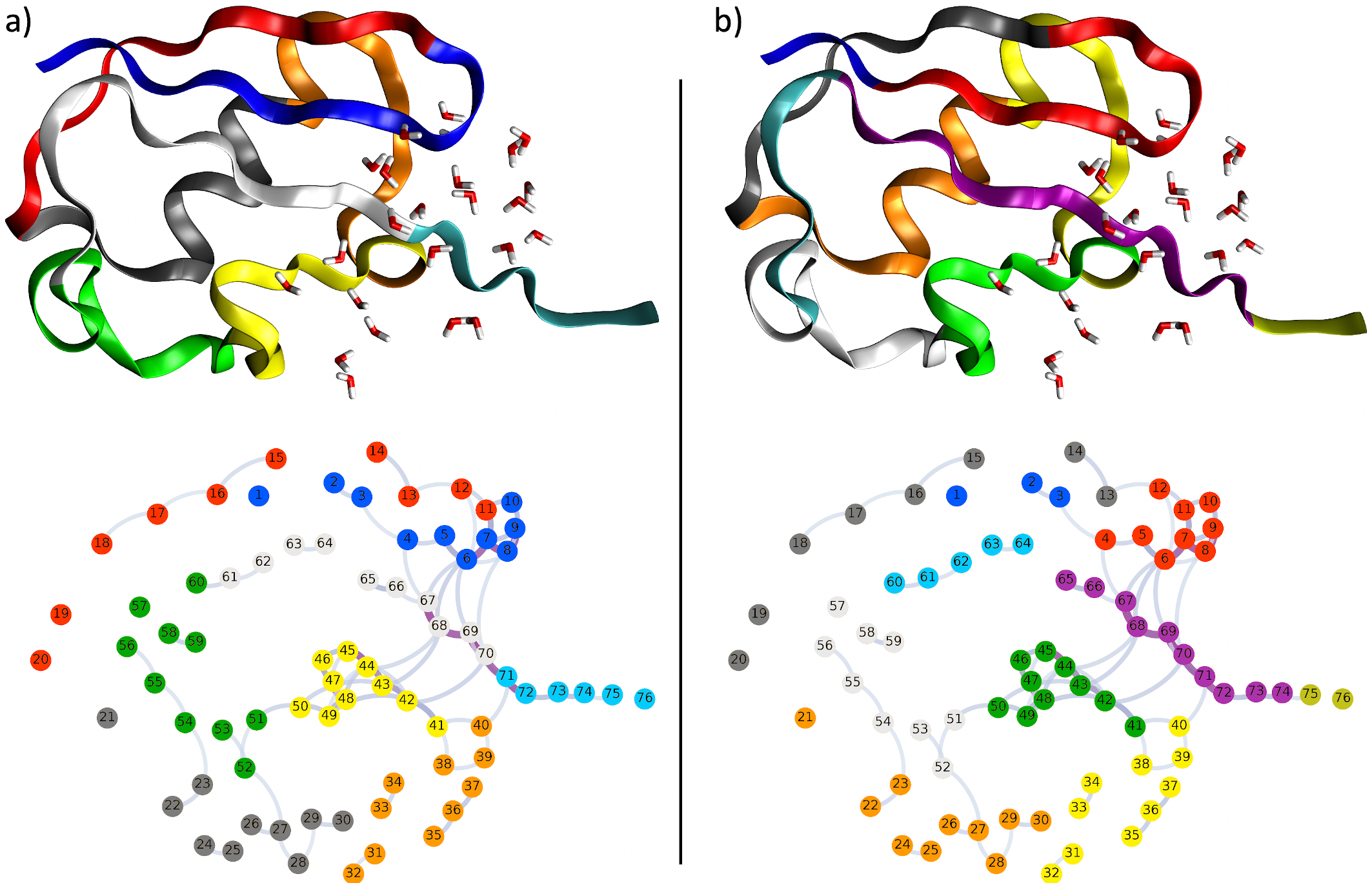}
	\end{center}

	\caption{Visualization of (a) the naïve partition and (b) the partition found by the DP algorithm for a maximum fragment size
	              of $n_\text{max} = 10$. The upper part shows a cartoon representation of the protein in which the individual fragments
	              are highlighted in different color. The water molecules defining the RoI are also shown. The lower part shows a 
	              representation of the corresponding graph, in which the nodes are shown as colored circles while the edge weights 
	              are represented by the thickness and color of the connecting lines (only edges with weights above a chosen threshold
	              are shown). \\[1ex]}
	\label{fig:ubi_gephi_graph_naïve}
\end{figure}

Fig.~\ref{fig:ubi_gephi_graph_naïve} takes a closer look at the fragmentation of ubiquitin for a maximum fragment size of $n_\text{max} = 10$. 
The upper part of Fig.~\ref{fig:ubi_gephi_graph_naïve}a shows the corresponding naïve partition in a cartoon representation of the protein 
structure, in which the colors indicate the different fragments, whereas the lower part visualizes the corresponding graphs.
The largest edge weights appear between amino acids that are closest to the RoI defined by the water molecules, in particular for the 
$\beta$-turn formed by amino acids 4 -- 12 and parts of the $\beta$-sheet (amino acids 65 -- 71 as well as amino acids 41 -- 49). For the
naïve partition, several cuts appear in these important regions, such as between amino acids 70 and 71 as well as between amino
acids 10 and 11. These cuts result in a large fragmentation error in the RoI.

Fig.~\ref{fig:ubi_gephi_graph_naïve}b shows the corresponding partition from the DP algorithm, for which the fragmentation error is
reduced significantly (see Fig.~\ref{fig:ubi_naïve_dp}d). Here, all imperfections of the naïve partition have been obliterated and the 
DP algorithm optimizes the partition such that no cuts appear too close to the RoI. The $\beta$-turn (amino acids 4--12) 
as well as the individual strands of the $\beta$-sheet are now kept intact. Thus, by minimizing the edge cut weight, the DP 
algorithm automatically finds a partition in which chemically meaningful subunits of the protein are kept within the same fragment.

\section{Application to Additional Test Cases}
\label{sec:proteins}

\begin{figure}
	\begin{center}
		\includegraphics[width=0.9\linewidth]{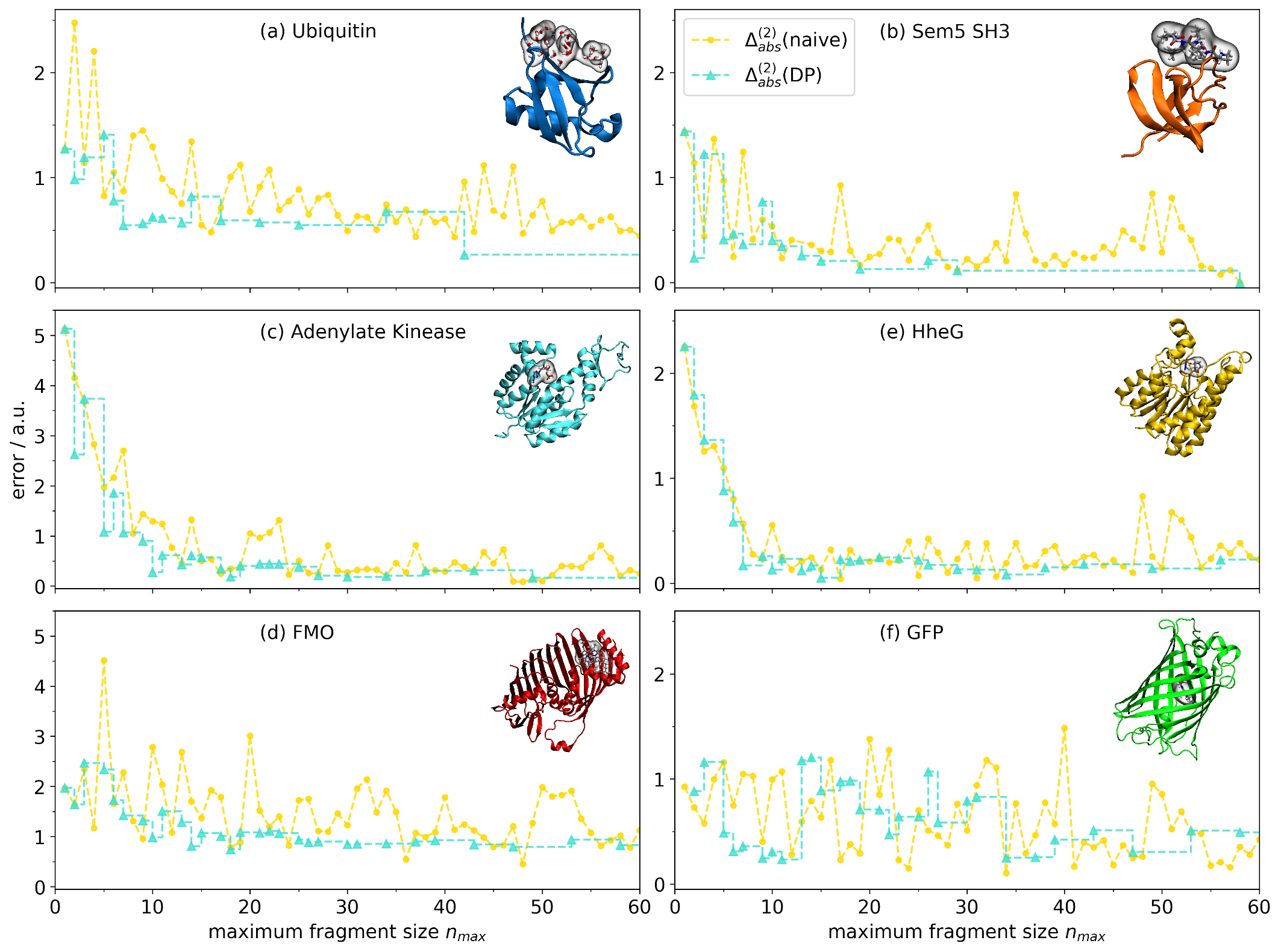}
	\end{center}
	
	\caption{Comparison of the two-body estimate for the absolute fragmentation error $\Delta_\text{abs}^{(2)}$ for the
	              naïve partition (yellow circles) and the partition obtained by the DP algorithm (cyan triangles) for increasing 
	              maximum number of amino acids per fragment $n_\text{max}$. (a) Ubiquitin; (b) Sem5 SH3 domain;
	              (c) adenylate kinease; (d) Fenna-Mathews-Olson complex (FMO); (e) halohydrin dehalogenase G (HheG); (f) green fluorescent protein (GFP).\\[1ex]}
	\label{fig:error_proteins}
\end{figure}

In the following, we apply the methodology established above to the further test cases shown in Fig.~\ref{fig:proteins_roi}. Since
for the larger proteins, a supermolecular calculation for the full protein is not easily possible, we only compare the error estimate
$\Delta_\text{abs}^{(2)}$, which has been shown to be a sufficiently accurate approximation of the absolute fragmentation error
$\Delta_\text{abs}$ for ubiquitin. Fig.~\ref{fig:error_proteins} compares $\Delta_\text{abs}^{(2)}$ for the naïve partition and for
the one obtained with the DP algorithm as function of the maximum number of amino acids per fragment $n_\text{max}$. For 
reference, the results for ubiquitin discussed above are shown in Fig.~\ref{fig:error_proteins}a.

As test case of protein--ligand interactions, we used the crystal structure of the Sem5 SH3 domain in a complex with a peptoid 
inhibitor \cite{nguyen_exploiting_1998} (PDB code: 3SEM \cite{nguyen_exploiting_1998}, see Fig.~\ref{fig:proteins_roi}b). The 
protein consists of 58 amino acids and this example was used earlier by Antony and Grimme to assess the accuracy of protein--ligand 
interaction energies calculated with the MFCC scheme \cite{antony_fully_2012}. Our results for Sem5 SH3 are shown in 
Fig.~\ref{fig:error_proteins}b. For the naïve partition, there are large fluctuations in $\Delta_\text{abs}^{(2)}$ when increasing 
the maximum fragment size. For instance, very large errors are obtained for $n_\text{max} = 7$ and for $n_\text{max} = 17$. 
Except for small maximum fragment sizes ($n_\text{max} = 1, 4$), the DP algorithm consistently reduces $\Delta_\text{abs}^{(2)}$ 
compared to the naïve partition and $\Delta_\text{abs}^{(2)}$ mostly shows a monotonic decrease when increasing the maximum 
fragment size. The only exception is $n_\text{max} = 9$, for which the DP algorithm yields a slightly higher error estimate than 
the naïve partition. 

The next two test cases are examples of enzymes, in which we consider the substrate binding pocket as the RoI. First, we consider
the crystal structure of adenylate kinase, which has 226 amino acids, with an adenosine monophosphate molecule in its binding 
pocket (PDB code: 2AK3 \cite{diederichs_refined_1991}, see Fig.~\ref{fig:proteins_roi}c), which is another example studied earlier by 
Antony and Grimme \cite{antony_fully_2012}. Second, we consider halohydrin dehalogenase G (HheG) \cite{koopmeiners_hheg_2017} 
(consisting of 255 amino acids) where the RoI is represented by a cyclohexene oxide, an azide and a water molecule in its substrate binding 
pocket (see Fig.~\ref{fig:proteins_roi}d). Here, we use a snapshot from the molecular dynamics simulations performed for this 
system in our earlier work \cite{solarczek_position_2019}. 

The error estimates for the naïve and the DP partitions for these two systems can be found in Fig.~\ref{fig:error_proteins}c and~d,
showing only maximum fragment sizes up to 60 amino acids. For both systems, $\Delta_\text{abs}^{(2)}$ rapidly increases when 
increasing the maximum fragment size from 1 to ca.~10. Starting at $n_\text{max} = 5$, it is consistently lower for the DP algorithm 
than for the naïve partition. While the naïve partition shows comparably large error estimates for some specific values of $n_\text{max}$
(e.g., for 20--24 for adenylate kinase and for 10, 48, and 51 for HheG), the partitions obtained by the DP algorithms do not exhibit 
such irregularities.

Our final two test cases are examples of photoactive proteins that contain embedded chromophores. First, we use the FMO protein 
(358 amino acids, see Fig.~\ref{fig:proteins_roi}e), for which 3-FDE calculations of the excitation energies of the 
embedded bacteriochlorophylls (BChl) have been performed earlier \cite{goez_modeling_2014}. Here, we used the equilibrated 
structure from Ref.~\citenum{goez_modeling_2014} and consider the BChl~1 chromophore as the RoI. 
Second, we also included GFP (238 amino acids, see Fig.~\ref{fig:proteins_roi}f), for which we used the available crystal structure
(PDB code: 2GX2). Here, the chromophore and the resulting ends of the main amino acid chain are capped with hydrogen atoms as shown in Sect.~\ref{sec:frag-methods}.
Note that in contrast to our four previous test cases, the RoI is fully embedded inside the protein for both FMO and GFP.

The results for GFP and FMO are presented in Fig.~\ref{fig:error_proteins}e and~f. For FMO, the DP algorithm shows an overall
improvement compared to the naïve partition, with only a few exceptions for values of $n_\text{max}$ at which the naïve partition
performs particularly well. However, for the naïve partition $\Delta_\text{abs}^{(2)}$ shows large fluctuations when increasing the
maximum fragment size, while a monotonic decrease is found in the DP partitions. On the other hand, for GFP the DP algorithm and
the naïve approach result in partitions with overall similar error estimates, and none of the two variants clearly performs better. In
GFP, the chromophore is inside the $\beta$-barrel structure at the center of the protein and most amino acids will thus be close 
to the RoI, Therefore, it is not easily possible to move the cuts between amino acids away from the RoI, which makes GFP a 
particularly difficult case for fragmentation methods, in which hardly any improvement over the naïve approach is possible.

\section{Conclusions and Outlook}
\label{sec:conclusion}

We have presented a methodology for systematically partitioning proteins for quantum-chemical fragmentation methods, which
determines the fragmentation such that for a given maximum fragment size the expected fragmentation error in local target
quantities is (approximately) minimized. To this end, we map the protein to a graph representation, in which the nodes represent 
amino acids and the edges are assigned weights that correspond to the fragmentation error that is expected when cutting 
this edge in the fragmentation. This makes it possible to apply graph partitioning algorithms provided by computer science to 
determine a near-optimal partition of the protein.

Here, we have chosen to minimize the error in the Coulomb potential of the protein in a specific RoI. This error, in particular the 
absolute fragmentation error $\Delta_\text{abs}$ defined here, should be closely correlated with the errors in protein--ligand 
interaction energies, in energy differences between different species in the active center of an enzyme, or in local spectroscopic 
properties of embedded chromophores. However, for specific applications in any of these areas, it might be necessary to further 
refine the definition of the relevant fragmentation error.

For assigning the edge weights in the graph representation of proteins, we have applied a two-body approximation, which has 
been shown to be sufficiently accurate for our purposes. However, in the current implementation the determination of the edge 
weights requires quantum-chemical calculations for each pair of amino acids and thus imposes a significant computational 
overhead, even if interactions between distant amino acids are neglected. To address this, we plan to develop more efficient 
schemes in which the edge weights are parametrized in terms of simpler descriptors, such as the distances between the amino 
acids as well as their distance from the RoI. Similarly, more sophisticated screening strategies could significantly reduce the 
number of pairs for which explicit quantum-chemical calculations are required. Work in these directions is currently in progress 
in our research groups.

For simplicity, we only considered a simplified MFCC scheme using hydrogen caps as quantum-chemical fragmentation method
in the present work. This way, interactions between the fragments and the caps can be neglected. This is generally not the case
when using fragmentation methods with overlapping fragments, for which the mapping to a graph representation needs to be
generalized by including additional nodes representing the caps, which will require using negative edge weights as well as 
additional constraints in the graph partitioning algorithms. As the present work demonstrates the usefulness of our graph-based
methodology, we plan to pursue such an extension in our future work.

Using a test set of six different proteins, we could show that with the simplified MFCC scheme, our graph-based partitioning 
scheme using the DP algorithm consistently reduces the expected fragmentation error compared to the commonly used naïve 
partitioning strategy for the same maximum fragment size (i.e., for a comparable computational effort). This is particularly true 
for maximum fragment sizes between 5 and 20 amino acids, which are most relevant for the application of quantum-chemical 
fragmentation methods. In most cases, for the optimized partitions obtained with the DP algorithm, the error is systematically 
reduced when increasing the maximum fragment size, while large oscillations of the expected error are found for the naïve 
partitions. The magnitude of improvement brought about by our graph-based partitioning scheme depends on the considered 
proteins as well as the choice of the RoI. While for cases with a well-localized RoI, a significant reduction of the fragmentation 
error is possible, we found that no or only little improvement with respect to the naïve partition for proteins  in which the RoI 
is at the center of the protein, such as GFP. 

In the present work, we have used the maximum number of amino acids per fragment as tunable parameter, which can be chosen
such that the computational effort for the quantum-chemical calculations of the fragments remains feasible. However, this 
computational effort depends of the number of basis functions (which will be roughly proportional to the number of atoms) 
and the different amino acids in a protein each contain different numbers of atoms. Thus, the maximum number of atoms per 
fragment would be a more meaningful parameter, but requires the modification of existing graph partitioning algorithms. We
plan to address this point in the future.

\section*{Acknowledgments}

M.W. and C.R.J. acknowledge funding from the Deutsche Forschungsgemeinschaft for the development of PyADF (Project Suresoft, 
JA 2329/7-1). M.v.L. and H.M. acknowledge funding from the Federal Ministry of Education and Research (Project WAVE, 01|H15004B).

\section*{Supporting Information Available}

Additional results for Sem5 SH3; additional results for ubiquitin without applying distance cut-offs.

\providecommand{\latin}[1]{#1}
\providecommand*\mcitethebibliography{\thebibliography}
\csname @ifundefined\endcsname{endmcitethebibliography}
  {\let\endmcitethebibliography\endthebibliography}{}

\end{document}